\documentclass[fleqn,usenatbib]{mnras}
\usepackage{newtxtext,newtxmath}

\usepackage[T1]{fontenc}
\usepackage{ae,aecompl}
\usepackage{graphicx}
\usepackage{xcolor}

\title{Kinematic and thermal signatures of the directly imaged protoplanet candidate around Elias 2-24}

\graphicspath{{./}{figures/}}

\author[Pinte et al.]{C. Pinte$^{1,2}$, I. Hammond$^{1}$, D. J. Price$^{1}$, V. Christiaens$^{1,3}$, S. M. Andrews$^{4}$, G. Chauvin$^{5,2}$, L. M. P\'erez$^{6,7}$,
\newauthor{S. Jorquera$^{6}$, H. Garg$^{1}$, B. J. Norfolk$^{8}$, J. Calcino$^{9}$}, M. Bonnefoy$^{2}$\\
$^{1}$School of Physics and Astronomy, Monash University, Vic 3800, Australia\\
$^{2}$Univ. Grenoble Alpes, CNRS, IPAG, F-38000 Grenoble, France\\
$^{3}$Space sciences, Technologies \& Astrophysics Research (STAR) Institute, Universit\'e de Li\`ege, All\'ee du Six Ao\^ut 19c, B-4000 Sart Tilman, Belgium \\
$^{4}$Centre for Astrophysics and Supercomputing (CAS), Swinburne University of Technology, Hawthorn, Victoria 3122, Australia \\
$^5$Université Côte d'Azur, Observatoire de la Côte d'Azur, CNRS, Laboratoire Lagrange, France\\
$^6$Departamento de Astronomía, Universidad de Chile, Camino El Observatorio 1515, Las Condes, Santiago, Chile\\
$^7$N\'ucleo Milenio de Formaci\'on Planetaria (NPF), Chile\\
$^8$Centre for Astrophysics and Supercomputing (CAS), Swinburne University of Technology, Hawthorn, Victoria 3122, Australia\\
$^9$Theoretical Division, Los Alamos National Laboratory, Los Alamos, NM 87545, USA
}

\defcitealias{Jorquera2021}{J21}

\pubyear{2022}

\begin{document}
\maketitle

\begin{abstract}
We report kinematic and thermal signatures associated with the directly imaged protoplanet candidate in the Elias 2-24 disc. Using the DSHARP ALMA observations of the $^{12}$CO J=2-1 line, we show that the disc kinematics are perturbed, with a detached CO emission spot at the location of the planet candidate and traces of spiral wakes, and also that the observed CO emission intensities require local heating.
While the foreground extinction hides the velocity channels associated with the planet, preventing a planet mass estimate,
the level of gas heating implied by the CO emission indicates the presence of a warm, embedded giant planet. Comparison with models show this could either be a $\gtrsim 5$M$_\mathrm{Jup}$, or a lower mass ( $\gtrsim 2$M$_\mathrm{Jup}$) but accreting proto-planet.
\end{abstract}


\label{firstpage}
\section{Introduction} \label{sec:intro}

\begin{figure*}
  \includegraphics[width=\hsize]{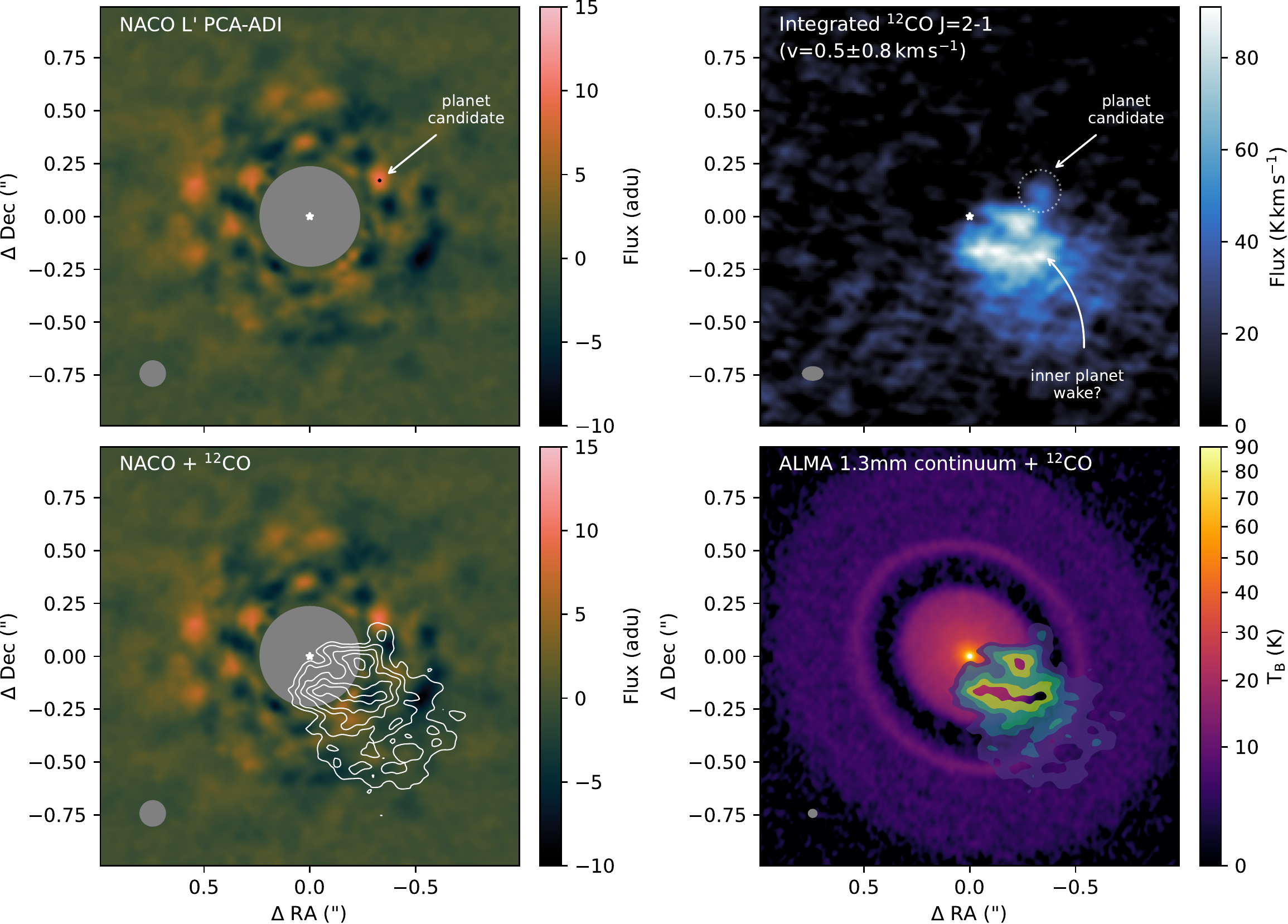}
  \vspace{-5mm}
  \caption{Kinematic counterpart of the protoplanet candidate detected around Elias2-24. \emph{Top left:} re-imaging of the NaCo L'-band PCA-ADI computed in concentric annuli with 20 principle components subtracted. The bright source discovered by \citet{Jorquera2021} is re-detected. \emph{Top right:}
    ALMA $^{12}$CO J=2-1 integrated line intensity between -0.375 and 1.375 km\,s$^{-1}$, revealing a CO spot detached from the isovelocity curve, as well a potential spiral arm.  \emph{Bottom:} NaCo image (\emph{left}) and ALMA band 6 continuum (\emph{right}) with integrated $^{12}$CO emission overlaid.}
  \label{fig:data}
\end{figure*}

The characterisation of young, still embedded, and potentially forming planets is in its infancy. Detecting them via direct imaging has proven more challenging than first envisaged, with the new generation of adaptive optics systems having yielded only two confirmed exoplanet detections within a protoplanetary disc
to date (e.g. \citealp{Keppler2018, Muller2018, Christiaens2019, Haffert2019},
see reviews by \citealp{BenistyPPVII} and \citealp{CurriePPVII}).
The high sensitivity of the Atacama Large Millimetre/submillimetre Array (ALMA) enabled the developement of a new planet detection method. By mapping of perturbations to the disc kinematics via high spectral resolution line imaging, ALMA can reveal the presence of otherwise hidden planets (e.g. \cite{Perez2015,Perez2018}, see review by \citealp{PintePPVII}). This led to the detection of a growing number of planet candidates \citep[e.g.][]{Pinte2018,Pinte2019,Pinte2020,Teague2018,Teague2021,Teague2022,Casassus2019,Casassus2022,Perez2020,Calcino2022,Izquierdo2022,Bae2022,Verrios2022,Garg2022}.

Despite these promising results, there is not yet any example of a detection with both disc kinematics and direct imaging. Such a detection would
constrain both the luminosity (via direct imaging) and the mass (via disc kinematics) of a newborn planet, thus distinguishing whether planets are born in a hot, cold, or warm start scenario \citep[e.g.][]{Marley2007,Spiegel2012}.
In this Letter we report such a detection in the disc of Elias 2-24 where \citet[][hereafter \citetalias{Jorquera2021}]{Jorquera2021} found evidence for a planet in images from the Nasmyth Adaptive Optics System Near-Infrared Imager and Spectrograph (NaCo) on the VLT.

\section{Observations and data reduction}
\label{sec:methods}

We re-analysed the NaCo Elias2-24 observations in $L'$-band from July 15, 2018 (\citetalias{Jorquera2021}), using
the same pre-processed Angular Differential Imaging (ADI) cube, stellar point spread function and parallactic angles acquired by \citetalias{Jorquera2021} and the same platescale and true north calibrations. After bad frame removal, the cube contained 414 frames with total on-source integration time of 82.8 minutes and full-width at half-maximum (FWHM) of 0.116". %4.28 pixels
After subtracting the median background, we post-processed the cube using median ADI, full-frame Principal Component Analysis PCA-ADI and PCA-ADI in concentric annuli using our NaCo data reduction pipeline \citep{Hammond2022}, based on the Vortex Image Processing package ({\sc vip},
\citealt{GomezGonzalez2017}). For PCA-ADI in concentric annuli, we used a 1-FWHM rotation angle threshold, a 1-FWHM inner mask and 12-pixel (0.326") wide annuli
to 50 principal components. We excluded the inner $\sim$0.22" of the post-processed images which contained residual speckle noise overlapping the saturated PSF core.

We used the publicly available $^{12}$CO data cube and measurement set from the Disc Substructures at High Angular Resolution Project (DSHARP; \citealt{Andrews2018}).
The limited signal-to-noise and $uv$ coverage of the data --- where the main goal was to image continuum at high spatial resolution --- can lead to image synthesis artefacts in CO emission potentially mimicking kinematic deviations. Continuum subtraction can also affect the line brightness and morphology of optically thick line emission when continuum and line intensity are comparable \citep{Boehler2017}. To test these effects, we followed \cite{Pinte2020} and imaged the data for a set of imaging parameters (robust parameter and Gaussian taper size), with and without continuum subtraction and the ``JvM'' correction \citep{Jorsater1995,Czekala2021}. For all imaging attempts, we used the auto-masking function of {\sc tclean}.
The planet signal we report below was detected in all resulting images. Hence, we only show the fiducial DSHARP cube (beam size 0.09"$\times$0.06" at PA=90$^\mathrm{o}$ with a rms of 1.4mJy/beam per channel, no JvM correction).

\begin{figure*}
  \includegraphics[width=\hsize]{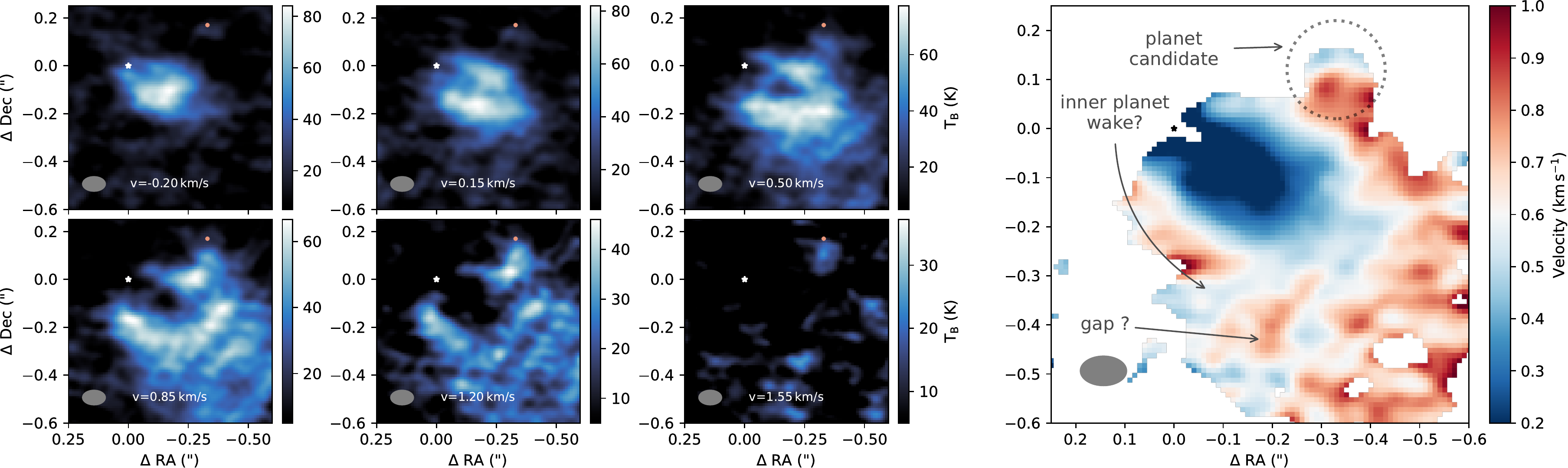}
  \vspace{-4mm}
  \caption{Velocity structures near the planet candidate. \emph{Left:} $^{12}$CO channel maps over which the detached CO spot is detected. Pink dots indicate the position of the NaCo source.
  Top left panel of Fig.~1 is the sum of these 6 channels. \emph{Right:} $^{12}$CO moment 1 over the same velocity range. The white contour shows a ``kink'' at the location of the planet candidate. A tentative gap structure and/or inner planet wake is also detected in velocity. \label{fig:channels}}
\end{figure*}

\begin{figure*}
  \includegraphics[width=\hsize]{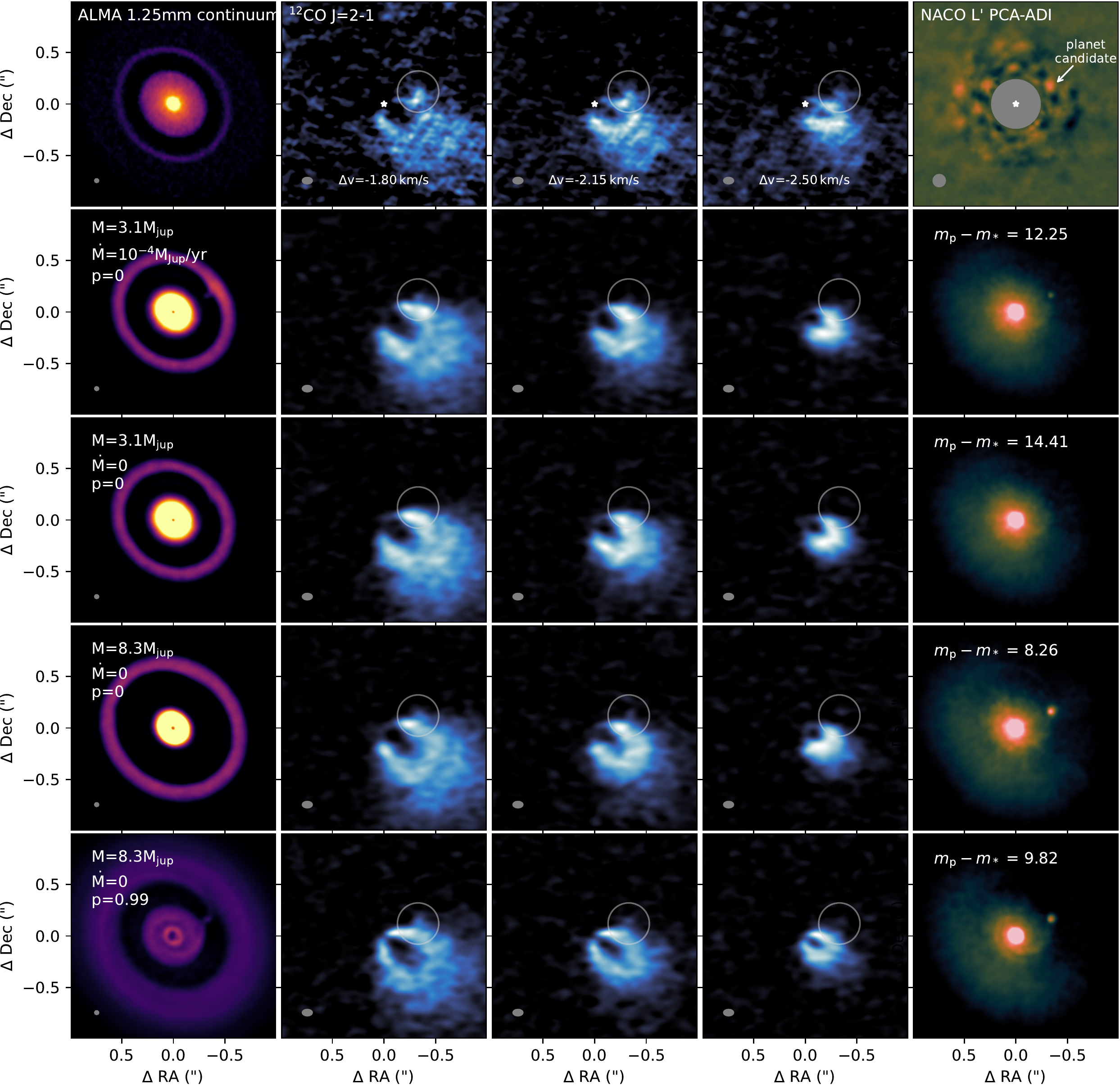}
  \vspace{-5mm}
  \caption{Comparison of ALMA and NaCo observations with synthetic maps with embedded planets of 3 and 8 M$_\mathrm{Jup}$ with various planet accretion rate and dust porosity. We adopted $v_\mathrm{syst} = 3$ km\,s$^{-1}$. The grey circles highlight the kinematic structures created by the planets. ALMA synthetic maps were convolved with a Gaussian beam and Hanning kernel to match the spatial and spectral resolution of the observations. The synthetic NaCo images were not ADI processed, in panel we indicate the magnitude difference between the planet and the central star.
    \label{fig:mcfost+phantom}}
\end{figure*}

\section{Results}
\label{sec:results}

We recover the \citetalias{Jorquera2021} detection in our re-reduction of their data using classical ADI and PCA-ADI (Fig.~\ref{fig:data}).  We estimated the flux and position of the source by minimizing the residuals after injecting a negative fake companion (NEGFC) using a simplex Nelder-Mead minimization function. We then sampled the parameter space around the minima using a Markov chain Monte Carlo (MCMC) with $\sim$4,500 steps and 120 walkers. We used the new log-probability function introduced by \cite{Christiaens2021} to take into account the effect of stellar speckles.
We obtained a separation of 0.37$\pm$0.04" ($\sim$51.8 au) and a PA of 298.4$\pm$4.6$^{\circ}$. Using the flux from the star in one FWHM aperture and flux from the candidate,
 we derived an $L'$ contrast of 8.78$^{+0.98}_{-0.48}$ mag.

The location is consistent within 1-$\sigma$ with \citetalias{Jorquera2021}'s reported separation of 0.411$\pm$0.008\,", PA=302.10$\pm$1.14\,deg, and $\Delta L'=8.81\pm0.12$\,mag.
The differences are due to high-pass filtering applied in \citetalias{Jorquera2021} but not here, and our uncertainties are larger as we used a MCMC to explore correlations between parameters.

The planet candidate is close to the Airy ring caused by the saturated star in the absence of a coronagraph. Our imaging shows additional compact signals at similar level to the planet candidate, in particular to the South-East of the star. These may be due to imperfect subtraction near the Airy ring and/or to ADI filtering of the brighter disc surface (this corresponds to the near side of the disc, where we expect forward scattering). We created two contrast curves by considering either the entire frame or a wedge spanning a PA range of
270-20\degr~to avoid the bright extended feature in the frame. We injected fake companions in the calibrated cube at varying radii to estimate the flux loss caused by post-processing.
The final contrast curve considers Student statistics at small separation as proposed in \citet{Mawet2014}, placing our detection at 4$\sigma$ in the wedge, or $\approx$ 2$\sigma$ in the full frame.

Fig.~\ref{fig:data}, top right shows a bright $^{12}$CO spot detached from the expected location of the emitting region along the isovelocity curve in the ALMA data, and located close to the NaCo $L'$ planet candidate. The morphology matches that predicted by \cite{Perez2015} for emission from a circumplanetary disc (CPD), and is similar to the CPD candidate detection in $^{13}$CO by \cite{Bae2022} in AS~209. The CO spot lies inside the dust gap observed in 1.3mm continuum emission (Fig.~\ref{fig:data}, bottom right).
We detect the CO spot in 6 channels  (imaged with a channel width of 350\,m\,s$^{-1}$) from -0.20 to 1.55\,km\,s$^{-1}$ (Fig.~\ref{fig:channels}), which corresponds to 3 spectral resolution elements (velocity resolution of 700\,m\,s$^{-1}$), with a significance of 4 to 10 times the rms. In individual channels, the spot is spatially unresolved, but its position moves slightly from channel to channel, resulting in a velocity gradient across the spot. We measure a separation of 0.36$\pm$0.04" and PA = 290$\pm$3$^{\circ}$ by fitting a Gaussian to the CO spot. Channels at velocities larger than 1.55\,km\,s$^{-1}$ are contaminated by the intervening cloud and/or envelope. The missing emission means that the measured position of the CO spot may be shifted from the actual planet location.

 At 1.55\,km\,s$^{-1}$, most of the disc emission has disappeared apart from the CO spot itself, indicating that CO spot is brighter than the surrounding disc. Contamination starts to impact the data at v$\approx$0.5\,km\,s$^{-1}$, visible as a decreasing brightness temperature in Fig.~\ref{fig:channels}. In particular, we do not detect any emission in  the channel where the isovelocity curve would correspond to the location of the planet candidate. This prevents us from detecting a potential velocity kink as seen in HD~163296 \citep{Pinte2018}, HD~97048 \citep{Pinte2019} or AS~209 \citep{Bae2022}. Nonetheless, a partial moment 1 map (\emph{i.e.} integrated over velocities between -0.20 and 1.55\,km\,s$^{-1}$) shows a significant shift in velocity at the location of the planet candidate, as illustrated by the white level in the right panel of Fig.~\ref{fig:channels}, which traces the isovelocity at 0.6\,km\,s$^{-1}$. We caution here that this does not allow for a measurement of the actual velocity shift at the location of the planet candidate, as some emission at the same location is missing due to contaminated channels, resulting in a skewed moment map.
The partial moment 1 map (right panel of Fig.~\ref{fig:channels}) also confirms the observed velocity gradient across the CO spot. Due to the missing emission in subsequent channels, it is not possible to determine if the velocity gradient is from the global Keplerian rotation of the disc or from additional velocity perturbations near the planet candidate. For the same reason, the extincted channels prevent us from measuring the local line width, but the detection of the CO spot in 6 channels indicates it is larger than in the surrounding disc. This is consistent with the prediction from an embedded planet (Fig. 3 in \citealp{Perez2015}).

\section{Models and disccussion}

To check if an embedded planet can explain the observed kinematic structures, we performed a series of 3D multigrain gas+dust simulations with the Smoothed Particle Hydrodynamics (SPH) code {\sc phantom} \citep{Price2018}, using $10^6$ particles and evolving the dust fraction with 11 grain sizes between 1\,$\mu$m and 1\,mm employing the one fluid formalism \citep{Laibe2014,Price2015,Hutchison2018}. Comparison with evolutionary tracks indicates a low stellar mass \citep[$\approx$ 0.8M$_\odot$,][]{Wilking2005,Andrews2010,Andrews2018}, but the high extinction (A$_V \approx 8.5$) makes this estimate uncertain. By matching isovelocity contours to the (few) non-extincted red and blue channels, we find a dynamical stellar mass close to $1.5$M$_\odot$, which we adopt here, and the systemic velocity is $\approx$ 3.0\,km\,s$^{-1}$. We set the accretion radius to 1\,au. We setup a disc with an initial gas mass of 0.01M$_\odot$ between 5 and 180\,au, a tapered surface density profile:
$$
\Sigma(r) = \Sigma_c \left( \frac{r}{r_c}\right)^{-\gamma} \exp\left( -\left( \frac{r}{r_c}\right)^{2-\gamma} \right)
$$
with $\gamma=-0.8$ and $r_c=120$\,au. We adopted a vertically isothermal equation of state with $H/R=0.1$ at $r=50$\,au and sound speed power-law index of -1/3. The density and temperature profiles impact the planet wake shape, which we cannot constrain here due to contamination, but not the kinematic signature near the planet, so we kept these values fixed. We set the shock viscosity $\alpha_{\rm av}=0.2$ to obtain a mean Shakura-Sunyaev viscosity of $5\times 10^{-3}$. This is above the lower bound of $\alpha_{\rm av}\approx0.1$ to resolve physical viscosity in {\sc phantom} \citep{Price2018}.

We embedded a planet at 65\,au with an initial mass of either 1,3,5,7 or 10\,M$_\mathrm{Jup}$, with accretion radius set to 0.125 times the Hill radius. We evolved the simulations for 100 planet orbits, by which time the planets reached a mass of 3.1, 5.8, 8.3, 10.5 and 12.5\,M$_\mathrm{Jup}$, and migrated to radii between 56 and 60\,au.

We post-processed the models with the radiative transfer code {\sc mcfost} \citep{Pinte2006,Pinte2009} to compute the dust temperature structure and synthetic continuum and CO maps, matching Voronoi cells to SPH particles. We assumed a distance of 140\,pc \citep{GaiaDR3} and 1Myr isochrones to set the star \citep{Siess2000} and planet \citep{Allard2012} luminosities and effective temperature, giving a radius of 3\,R$_\odot$, and $T_{\rm eff} = 4600$\,K for the central star. These values differ slightly from the published photometric estimates (4.5\,R$_\odot$, and $4250$\,K), but those estimates suffer from high extinction and the morphology of the emission does not depend on the stellar parameter. For the planets, we obtained radii of 0.19, 0.22, 0.26, 0.29 and 0.32\,R$_\odot$ and $T_{\rm eff} = 1530$, 1960, 2150, 2250 and 2350\,K.
Because the accretion rate on the planet varied during the simulations, we used it as an additional free parameter in the radiative transfer, assuming the accretion luminosity is radiated from the planet surface (see \citealp{Borchert2022b} for details).

We assumed astrosilicate grains \citep{Weingartner2001} with sizes following
$\mathrm{d}n(a) \propto a^{-3.5}\mathrm{d}a$ between 0.03 to
1000$\mu$m, a gas-to-dust ratio of 100, and computed the dust optical properties using Mie theory. Following \cite{Pinte2009}, we allowed for fluffy grains by assigning a larger Stokes number for a given grain size. We set $T_\mathrm{gas} = T_\mathrm{dust}$, and assumed that the CO 2-1 transition is at local thermodynamic equilibrium. We set the CO abundance following the prescription in Appendix B of \cite{Pinte2018} to account for freeze-out where T $< 20$\,K, as well as photo-dissociation and photo-desorption. We set the turbulent velocity to zero.

Figure~\ref{fig:mcfost+phantom} shows the predicted emission for a selection of our model grid. A 3 M$_\mathrm{Jup}$ planet can produce a detached CO spot that resembles the observations but only if the planet is accreting at high rate ($10^{-4}$M$_\mathrm{Jup}$/yr; second row). Without accretion on the planet, no CO spot is visible (third row), and the planet is almost completely hidden by the disc in scattered light.
The high accretion rate model results in local heating of the outer dust ring in our model, which is not observed in the 1.25mm continuum data. Alternatively, a more massive planet ($\approx$ 8M$_\mathrm{Jup}$) can reproduce the observed CO spot without significant accretion (fourth row). With compact dust grains, this results in a wider dust gap in the model than in the 1.25mm observations. To simultaneously reproduce the dust gap and CO spot with such a massive planet, our model requires porous grains (bottom row).
In that case a ``circumplanetary disc" (CPD) is visible in the continuum emission, contrary to the data \citep{Andrews2021}. We note however that the CO spot does not trace a possible CPD, but kinematic structures high above the midplane: performing the same radiative transfer simulations after removing the Hill sphere results in indistinguishable channel maps.

Rescaling for a central mass of 1.5M$_\odot$, \cite{Cieza2017}, \cite{Dipierro2018} and \cite{Zhang2018} estimated from the depth and width of the dust gap a planet mass of 1.5 to 12, $\approx$1, and 0.4 to 3.3 M$_\mathrm{Jup}$ respectively. The lower bound of these estimates is about 10 times lower than our minimum mass of 3\,M$_\mathrm{Jup}$. We found a similar discrepancy for HD~97048 \citep{Pinte2019} and the other DSHARP sources with tentative velocity kinks \citep{Pinte2020}, where we suggested that the dust grains dominating the sub-millimetre emission have Stokes numbers of a few $10^{-2}$, like in our highly porous model (Fig.~\ref{fig:mcfost+phantom}, bottom row).

\citetalias{Jorquera2021} found a mass between 0.5 and 5 M$_\mathrm{Jup}$ for the planet by comparing the $L'$ contrast to predictions of masses and accretion rates from \cite{Zhu2015} for embedded planets with circumplanetary discs. Our models suggest the planet may be more massive. All our models, including the ones with a 8 M$_\mathrm{Jup}$ planet, result in a contrast higher or consistent with the observations, as the intrinsic emission from the planet and its surrounding environment are significantly extincted by the disc. The observed luminosity of the planet depends on properties including its accretion rate, its potential CPD \citep[e.g.][]{Szulagyi2019}, but also crucially on the circumstellar disc optical depth at the planet location. We thus cannot exclude the 3M$_\mathrm{Jup}$ models, as a lower infrared dust opacity would make the planet more visible.

None of our models reproduce all the observations. They should only be seen as a guide to try to understand the emission in the sub-millimetre and infrared continuum, and in CO lines. Our main limitation is the heavy cloud contamination which hides most of the disc --- in particular the channel where the disc isovelocity curve overlaps the planet location, where we expect a velocity kink ($v \approx$ 1.8 km/s). Hence it is not possible to estimate a robust planet mass from kinematics. Deeper observations with less contamination, \emph{i.e.} higher-J $^{12}$CO or in less abundant isotopes are required. Similarly, \cite{Cleeves2015} predicted that embedded giant planets should have observable ``chemical" signatures, in tracers like HCN or CS.
Our model requires the planet to heat its surrounding environment either because it is massive and hot or because it is accreting at a high rate. Elias 2-24 is a prime target to search for such a signature.

\section{Conclusions}
\label{sec:conclusions}

\begin{enumerate}
\item We detect a $^{12}$CO counterpart of the planet candidate presented by \citetalias{Jorquera2021}, and we recover the companion with independent reduction of the NaCo data. If confirmed this would be the first simultaneous planet detection via disc kinematics and direct imaging.
\item The  CO data shows a velocity shift, broadened linewidth, and increased disc temperature at the location of the planet candidate, all of which are predicted by models of planet-disc interactions.
\item These kinematic structures can be explained by a $\approx 8$M$_\mathrm{Jup}$ planet or a 3M$_\mathrm{Jup}$ accreting planet.
\item Cloud contamination prevents an accurate planet mass estimation. Higher-J and/or less abundant isotopologues/molecules can alleviate this, and together with the measured $L'$ flux, or improved data from \emph{e.g.} VLT/ERIS, can provide the first simultaneous dynamical and luminosity constraints on a forming planet candidate.
\end{enumerate}

\section*{Data and software avaibility}
Raw data sets are publicly available on the ALMA and ESO archives. Calibrated ALMA data sets are available at \url{alma-dl.mtk.nao.ac.jp/ftp/alma/lp/DSHARP}. Calibrated NACO data and synthetic models are available at \url{10.6084/m9.figshare.21915000}.
Our NaCo pipeline is available at: \url{github.com/ IainHammond/NaCo\_pipeline}. {\sc phantom} and {\sc mcfost} are publicly available at \url{github.com/danieljprice/phantom} and \url{github.com/cpinte/mcfost}. The scripts used to make the figures are available at \url{https://github.com/cpinte/elias2-24}.

\section*{Acknowledgments}
We made use of ALMA data ADS/JAO.ALMA\#2016.1.00484.L.
ALMA is a partnership of ESO (representing
its member states), NSF (USA) and NINS (Japan), together with NRC (Canada), MOST and ASIAA (Taiwan), and KASI (Republic of Korea), in cooperation with Chile. The Joint ALMA Observatory is operated by ESO, AUI/NRAO and
NAOJ. The National Radio Astronomy Observatory is a
facility of the National Science Foundation operated under
agreement with Associated Universities, Inc.
Simulations were performed on ozSTAR, funded by Swinburne University and the Australian Government.
C.P., V.C. and D.J.P. acknowledge Australian Research Council funding  via FT170100040, DP18010423 and DP220103767.
L.P. and S.J. gratefully acknowledge support from ANID BASAL project FB210003, ANID -- Millennium Science Initiative Program -- NCN19\_171, and ANID FONDECYT  1221442.

\bibliography{elias24}{}

\begin{thebibliography}{}
\makeatletter
\relax
\def\mn@urlcharsother{\let\do\@makeother \do\$\do\&\do\#\do\^\do\_\do\%\do\~}
\def\mn@doi{\begingroup\mn@urlcharsother \@ifnextchar [ {\mn@doi@}
  {\mn@doi@[]}}
\def\mn@doi@[#1]#2{\def\@tempa{#1}\ifx\@tempa\@empty \href
  {http://dx.doi.org/#2} {doi:#2}\else \href {http://dx.doi.org/#2} {#1}\fi
  \endgroup}
\def\mn@eprint#1#2{\mn@eprint@#1:#2::\@nil}
\def\mn@eprint@arXiv#1{\href {http://arxiv.org/abs/#1} {{\tt arXiv:#1}}}
\def\mn@eprint@dblp#1{\href {http://dblp.uni-trier.de/rec/bibtex/#1.xml}
  {dblp:#1}}
\def\mn@eprint@#1:#2:#3:#4\@nil{\def\@tempa {#1}\def\@tempb {#2}\def\@tempc
  {#3}\ifx \@tempc \@empty \let \@tempc \@tempb \let \@tempb \@tempa \fi \ifx
  \@tempb \@empty \def\@tempb {arXiv}\fi \@ifundefined
  {mn@eprint@\@tempb}{\@tempb:\@tempc}{\expandafter \expandafter \csname
  mn@eprint@\@tempb\endcsname \expandafter{\@tempc}}}

\bibitem[\protect\citeauthoryear{{Allard}, {Homeier}  \& {Freytag}}{{Allard}
  et~al.}{2012}]{Allard2012}
{Allard} F.,  {Homeier} D.,   {Freytag} B.,  2012, \mn@doi [Phil. Trans. Roy.
  Soc. A] {10.1098/rsta.2011.0269}, \href
  {https://ui.adsabs.harvard.edu/abs/2012RSPTA.370.2765A} {370, 2765}

\bibitem[\protect\citeauthoryear{{Andrews} et~al.}{{Andrews}
  et~al.}{2010}]{Andrews2010}
{Andrews} S.~M.,  et~al., 2010, \mn@doi [\apj] {10.1088/0004-637X/723/2/1241},
  \href {https://ui.adsabs.harvard.edu/abs/2010ApJ...723.1241A} {723, 1241}

\bibitem[\protect\citeauthoryear{{Andrews} et~al.,}{{Andrews}
  et~al.}{2018}]{Andrews2018}
{Andrews} S.~M.,  et~al., 2018, \mn@doi [\apjl] {10.3847/2041-8213/aaf741},
  \href {https://ui.adsabs.harvard.edu/abs/2018ApJ...869L..41A} {869, L41}

\bibitem[\protect\citeauthoryear{{Andrews} et~al.,}{{Andrews}
  et~al.}{2021}]{Andrews2021}
{Andrews} S.~M.,  et~al., 2021, \mn@doi [\apj] {10.3847/1538-4357/ac00b9},
  \href {https://ui.adsabs.harvard.edu/abs/2021ApJ...916...51A} {916, 51}

\bibitem[\protect\citeauthoryear{{Bae} et~al.,}{{Bae} et~al.}{2022}]{Bae2022}
{Bae} J.,  et~al., 2022, \mn@doi [\apjl] {10.3847/2041-8213/ac7fa3}, \href
  {https://ui.adsabs.harvard.edu/abs/2022ApJ...934L..20B} {934, L20}

\bibitem[\protect\citeauthoryear{{Benisty} et~al.,}{{Benisty}
  et~al.}{2022}]{BenistyPPVII}
{Benisty} M.,  et~al., 2022, arXiv e-prints, \href
  {https://ui.adsabs.harvard.edu/abs/2022arXiv220309991B} {p. arXiv:2203.09991}

\bibitem[\protect\citeauthoryear{{Boehler} et~al.}{{Boehler}
  et~al.}{2017}]{Boehler2017}
{Boehler} Y.,  et~al., 2017, \mn@doi [\apj] {10.3847/1538-4357/aa696c}, \href
  {https://ui.adsabs.harvard.edu/abs/2017ApJ...840...60B} {840, 60}

\bibitem[\protect\citeauthoryear{{Borchert}, {Price}, {Pinte}  \&
  {Cuello}}{{Borchert} et~al.}{2022}]{Borchert2022b}
{Borchert} E. M.~A.,  {Price} D.~J.,  {Pinte} C.,   {Cuello} N.,  2022, \mn@doi
  [\mnras] {10.1093/mnras/stac2872}, \href
  {https://ui.adsabs.harvard.edu/abs/2022MNRAS.tmp.2676B} {}

\bibitem[\protect\citeauthoryear{{Calcino} et~al.}{{Calcino}
  et~al.}{2022}]{Calcino2022}
{Calcino} J.,  et~al., 2022, \mn@doi [\apjl] {10.3847/2041-8213/ac64a7}, \href
  {https://ui.adsabs.harvard.edu/abs/2022ApJ...929L..25C} {929, L25}

\bibitem[\protect\citeauthoryear{{Casassus} \& {P{\'e}rez}}{{Casassus} \&
  {P{\'e}rez}}{2019}]{Casassus2019}
{Casassus} S.,  {P{\'e}rez} S.,  2019, \mn@doi [\apjl]
  {10.3847/2041-8213/ab4425}, \href
  {https://ui.adsabs.harvard.edu/abs/2019ApJ...883L..41C} {883, L41}

\bibitem[\protect\citeauthoryear{{Casassus}, {C{\'a}rcamo}, {Hales}, {Weber}
  \& {Dent}}{{Casassus} et~al.}{2022}]{Casassus2022}
{Casassus} S.,  {C{\'a}rcamo} M.,  {Hales} A.,  {Weber} P.,   {Dent} B.,  2022,
  \mn@doi [\apjl] {10.3847/2041-8213/ac75e8}, \href
  {https://ui.adsabs.harvard.edu/abs/2022ApJ...933L...4C} {933, L4}

\bibitem[\protect\citeauthoryear{{Christiaens} et~al.,}{{Christiaens}
  et~al.}{2019}]{Christiaens2019}
{Christiaens} V.,  et~al., 2019, \mn@doi [\mnras] {10.1093/mnras/stz1232},
  \href {https://ui.adsabs.harvard.edu/abs/2019MNRAS.486.5819C} {486, 5819}

\bibitem[\protect\citeauthoryear{{Christiaens} et~al.,}{{Christiaens}
  et~al.}{2021}]{Christiaens2021}
{Christiaens} V.,  et~al., 2021, \mn@doi [\mnras] {10.1093/mnras/stab480},
  \href {https://ui.adsabs.harvard.edu/abs/2021MNRAS.502.6117C} {502, 6117}

\bibitem[\protect\citeauthoryear{{Cieza} et~al.,}{{Cieza}
  et~al.}{2017}]{Cieza2017}
{Cieza} L.~A.,  et~al., 2017, \mn@doi [\apjl] {10.3847/2041-8213/aa9b7b}, \href
  {https://ui.adsabs.harvard.edu/abs/2017ApJ...851L..23C} {851, L23}

\bibitem[\protect\citeauthoryear{{Cleeves}, {Bergin}  \& {Harries}}{{Cleeves}
  et~al.}{2015}]{Cleeves2015}
{Cleeves} L.~I.,  {Bergin} E.~A.,   {Harries} T.~J.,  2015, \mn@doi [\apj]
  {10.1088/0004-637X/807/1/2}, \href
  {https://ui.adsabs.harvard.edu/abs/2015ApJ...807....2C} {807, 2}

\bibitem[\protect\citeauthoryear{{Currie} et~al.}{{Currie}
  et~al.}{2022}]{CurriePPVII}
{Currie} T.,  et~al., 2022, arXiv e-prints, \href
  {https://ui.adsabs.harvard.edu/abs/2022arXiv220505696C} {p. arXiv:2205.05696}

\bibitem[\protect\citeauthoryear{{Czekala} et~al.,}{{Czekala}
  et~al.}{2021}]{Czekala2021}
{Czekala} I.,  et~al., 2021, \mn@doi [\apjs] {10.3847/1538-4365/ac1430}, \href
  {https://ui.adsabs.harvard.edu/abs/2021ApJS..257....2C} {257, 2}

\bibitem[\protect\citeauthoryear{{Dipierro} et~al.,}{{Dipierro}
  et~al.}{2018}]{Dipierro2018}
{Dipierro} G.,  et~al., 2018, \mn@doi [\mnras] {10.1093/mnras/sty181}, \href
  {https://ui.adsabs.harvard.edu/abs/2018MNRAS.475.5296D} {475, 5296}

\bibitem[\protect\citeauthoryear{{Gaia Collaboration} et~al.,}{{Gaia
  Collaboration} et~al.}{2021}]{GaiaDR3}
{Gaia Collaboration} et~al., 2021, \mn@doi [\aap]
  {10.1051/0004-6361/202039657}, \href
  {https://ui.adsabs.harvard.edu/abs/2021A&A...649A...1G} {649, A1}

\bibitem[\protect\citeauthoryear{{Garg} et~al.,}{{Garg}
  et~al.}{2022}]{Garg2022}
{Garg} H.,  et~al., 2022, arXiv e-prints, \href
  {https://ui.adsabs.harvard.edu/abs/2022arXiv221010248G} {p. arXiv:2210.10248}

\bibitem[\protect\citeauthoryear{{Gomez Gonzalez} et~al.,}{{Gomez Gonzalez}
  et~al.}{2017}]{GomezGonzalez2017}
{Gomez Gonzalez} C.~A.,  et~al., 2017, \mn@doi [\aj]
  {10.3847/1538-3881/aa73d7}, \href
  {https://ui.adsabs.harvard.edu/abs/2017AJ....154....7G} {154, 7}

\bibitem[\protect\citeauthoryear{{Haffert} et~al.}{{Haffert}
  et~al.}{2019}]{Haffert2019}
{Haffert} S.~Y.,  et~al., 2019, \mn@doi [Nature Astronomy]
  {10.1038/s41550-019-0780-5}, \href
  {https://ui.adsabs.harvard.edu/abs/2019NatAs...3..749H} {3, 749}

\bibitem[\protect\citeauthoryear{{Hammond} et~al.,}{{Hammond}
  et~al.}{2022}]{Hammond2022}
{Hammond} I.,  et~al., 2022, \mn@doi [\mnras] {10.1093/mnras/stac2119}, \href
  {https://ui.adsabs.harvard.edu/abs/2022MNRAS.515.6109H} {515, 6109}

\bibitem[\protect\citeauthoryear{{Hutchison}, {Price}  \& {Laibe}}{{Hutchison}
  et~al.}{2018}]{Hutchison2018}
{Hutchison} M.,  {Price} D.~J.,   {Laibe} G.,  2018, \mn@doi [\mnras]
  {10.1093/mnras/sty367}, \href
  {https://ui.adsabs.harvard.edu/abs/2018MNRAS.476.2186H} {476, 2186}

\bibitem[\protect\citeauthoryear{{Izquierdo} et~al.}{{Izquierdo}
  et~al.}{2022}]{Izquierdo2022}
{Izquierdo} A.~F.,  et~al., 2022, \mn@doi [\apj] {10.3847/1538-4357/ac474d},
  \href {https://ui.adsabs.harvard.edu/abs/2022ApJ...928....2I} {928, 2}

\bibitem[\protect\citeauthoryear{{Jorquera} et~al.,}{{Jorquera}
  et~al.}{2021}]{Jorquera2021}
{Jorquera} S.,  et~al., 2021, \mn@doi [\aj] {10.3847/1538-3881/abd40d}, \href
  {https://ui.adsabs.harvard.edu/abs/2021AJ....161..146J} {161, 146}

\bibitem[\protect\citeauthoryear{{Jorsater} \& {van Moorsel}}{{Jorsater} \&
  {van Moorsel}}{1995}]{Jorsater1995}
{Jorsater} S.,  {van Moorsel} G.~A.,  1995, \mn@doi [\aj] {10.1086/117668},
  \href {https://ui.adsabs.harvard.edu/abs/1995AJ....110.2037J} {110, 2037}

\bibitem[\protect\citeauthoryear{{Keppler} et~al.,}{{Keppler}
  et~al.}{2018}]{Keppler2018}
{Keppler} M.,  et~al., 2018, \mn@doi [\aap] {10.1051/0004-6361/201832957},
  \href {https://ui.adsabs.harvard.edu/abs/2018A&A...617A..44K} {617, A44}

\bibitem[\protect\citeauthoryear{{Laibe} \& {Price}}{{Laibe} \&
  {Price}}{2014}]{Laibe2014}
{Laibe} G.,  {Price} D.~J.,  2014, \mn@doi [\mnras] {10.1093/mnras/stu355},
  \href {https://ui.adsabs.harvard.edu/abs/2014MNRAS.440.2136L} {440, 2136}

\bibitem[\protect\citeauthoryear{{Marley} et~al.}{{Marley}
  et~al.}{2007}]{Marley2007}
{Marley} M.~S.,  et~al., 2007, \mn@doi [\apj] {10.1086/509759}, \href
  {https://ui.adsabs.harvard.edu/abs/2007ApJ...655..541M} {655, 541}

\bibitem[\protect\citeauthoryear{{Mawet} et~al.,}{{Mawet}
  et~al.}{2014}]{Mawet2014}
{Mawet} D.,  et~al., 2014, \mn@doi [\apj] {10.1088/0004-637X/792/2/97}, \href
  {https://ui.adsabs.harvard.edu/abs/2014ApJ...792...97M} {792, 97}

\bibitem[\protect\citeauthoryear{{M{\"u}ller} et~al.,}{{M{\"u}ller}
  et~al.}{2018}]{Muller2018}
{M{\"u}ller} A.,  et~al., 2018, \aap, 617, L2

\bibitem[\protect\citeauthoryear{{Perez} et~al.}{{Perez}
  et~al.}{2015}]{Perez2015}
{Perez} S.,  et~al., 2015, \mn@doi [\apjl] {10.1088/2041-8205/811/1/L5}, \href
  {https://ui.adsabs.harvard.edu/abs/2015ApJ...811L...5P} {811, L5}

\bibitem[\protect\citeauthoryear{{P{\'e}rez}, {Casassus}  \&
  {Ben{\'\i}tez-Llambay}}{{P{\'e}rez} et~al.}{2018}]{Perez2018}
{P{\'e}rez} S.,  {Casassus} S.,   {Ben{\'\i}tez-Llambay} P.,  2018, \mn@doi
  [\mnras] {10.1093/mnrasl/sly109}, \href
  {https://ui.adsabs.harvard.edu/abs/2018MNRAS.480L..12P} {480, L12}

\bibitem[\protect\citeauthoryear{{P{\'e}rez} et~al.,}{{P{\'e}rez}
  et~al.}{2020}]{Perez2020}
{P{\'e}rez} S.,  et~al., 2020, \mn@doi [\apjl] {10.3847/2041-8213/ab6b2b},
  \href {https://ui.adsabs.harvard.edu/abs/2020ApJ...889L..24P} {889, L24}

\bibitem[\protect\citeauthoryear{{Pinte}, {M{\'e}nard}, {Duch{\^e}ne}  \&
  {Bastien}}{{Pinte} et~al.}{2006}]{Pinte2006}
{Pinte} C.,  {M{\'e}nard} F.,  {Duch{\^e}ne} G.,   {Bastien} P.,  2006, \mn@doi
  [\aap] {10.1051/0004-6361:20053275}, \href
  {https://ui.adsabs.harvard.edu/abs/2006A&A...459..797P} {459, 797}

\bibitem[\protect\citeauthoryear{{Pinte} et~al.}{{Pinte}
  et~al.}{2009}]{Pinte2009}
{Pinte} C.,  et~al., 2009, \mn@doi [\aap] {10.1051/0004-6361/200811555}, \href
  {https://ui.adsabs.harvard.edu/abs/2009A&A...498..967P} {498, 967}

\bibitem[\protect\citeauthoryear{{Pinte} et~al.,}{{Pinte}
  et~al.}{2018}]{Pinte2018}
{Pinte} C.,  et~al., 2018, \mn@doi [\apjl] {10.3847/2041-8213/aac6dc}, \href
  {https://ui.adsabs.harvard.edu/abs/2018ApJ...860L..13P} {860, L13}

\bibitem[\protect\citeauthoryear{{Pinte} et~al.,}{{Pinte}
  et~al.}{2019}]{Pinte2019}
{Pinte} C.,  et~al., 2019, \mn@doi [Nature Astronomy]
  {10.1038/s41550-019-0852-6}, \href
  {https://ui.adsabs.harvard.edu/abs/2019NatAs...3.1109P} {3, 1109}

\bibitem[\protect\citeauthoryear{{Pinte} et~al.,}{{Pinte}
  et~al.}{2020}]{Pinte2020}
{Pinte} C.,  et~al., 2020, \mn@doi [\apjl] {10.3847/2041-8213/ab6dda}, \href
  {https://ui.adsabs.harvard.edu/abs/2020ApJ...890L...9P} {890, L9}

\bibitem[\protect\citeauthoryear{{Pinte} et~al.}{{Pinte}
  et~al.}{2022}]{PintePPVII}
{Pinte} C.,  et~al., 2022, arXiv e-prints, \href
  {https://ui.adsabs.harvard.edu/abs/2022arXiv220309528P} {p. arXiv:2203.09528}

\bibitem[\protect\citeauthoryear{{Price} \& {Laibe}}{{Price} \&
  {Laibe}}{2015}]{Price2015}
{Price} D.~J.,  {Laibe} G.,  2015, \mn@doi [\mnras] {10.1093/mnras/stv996},
  \href {https://ui.adsabs.harvard.edu/abs/2015MNRAS.451..813P} {451, 813}

\bibitem[\protect\citeauthoryear{{Price} et~al.,}{{Price}
  et~al.}{2018}]{Price2018}
{Price} D.~J.,  et~al., 2018, \mn@doi [\pasa] {10.1017/pasa.2018.25}, \href
  {https://ui.adsabs.harvard.edu/abs/2018PASA...35...31P} {35, e031}

\bibitem[\protect\citeauthoryear{{Siess}, {Dufour}  \& {Forestini}}{{Siess}
  et~al.}{2000}]{Siess2000}
{Siess} L.,  {Dufour} E.,   {Forestini} M.,  2000, \aap, \href
  {https://ui.adsabs.harvard.edu/abs/2000A&A...358..593S} {358, 593}

\bibitem[\protect\citeauthoryear{{Spiegel} \& {Burrows}}{{Spiegel} \&
  {Burrows}}{2012}]{Spiegel2012}
{Spiegel} D.~S.,  {Burrows} A.,  2012, \mn@doi [\apj]
  {10.1088/0004-637X/745/2/174}, \href
  {https://ui.adsabs.harvard.edu/abs/2012ApJ...745..174S} {745, 174}

\bibitem[\protect\citeauthoryear{{Szul{\'a}gyi}, {Dullemond}, {Pohl}  \&
  {Quanz}}{{Szul{\'a}gyi} et~al.}{2019}]{Szulagyi2019}
{Szul{\'a}gyi} J.,  {Dullemond} C.~P.,  {Pohl} A.,   {Quanz} S.~P.,  2019,
  \mn@doi [\mnras] {10.1093/mnras/stz1326}, \href
  {https://ui.adsabs.harvard.edu/abs/2019MNRAS.487.1248S} {487, 1248}

\bibitem[\protect\citeauthoryear{{Teague} et~al.}{{Teague}
  et~al.}{2018}]{Teague2018}
{Teague} R.,  et~al., 2018, \mn@doi [\apjl] {10.3847/2041-8213/aac6d7}, \href
  {https://ui.adsabs.harvard.edu/abs/2018ApJ...860L..12T} {860, L12}

\bibitem[\protect\citeauthoryear{{Teague} et~al.}{{Teague}
  et~al.}{2021}]{Teague2021}
{Teague} R.,  et~al., 2021, \mn@doi [\apjs] {10.3847/1538-4365/ac1438}, \href
  {https://ui.adsabs.harvard.edu/abs/2021ApJS..257...18T} {257, 18}

\bibitem[\protect\citeauthoryear{{Teague} et~al.}{{Teague}
  et~al.}{2022}]{Teague2022}
{Teague} R.,  et~al., 2022, \mn@doi [\apj] {10.3847/1538-4357/ac88ca}, \href
  {https://ui.adsabs.harvard.edu/abs/2022ApJ...936..163T} {936, 163}

\bibitem[\protect\citeauthoryear{{Verrios}, {Price}, {Pinte}, {Hilder}  \&
  {Calcino}}{{Verrios} et~al.}{2022}]{Verrios2022}
{Verrios} H.~J.,  {Price} D.~J.,  {Pinte} C.,  {Hilder} T.,   {Calcino} J.,
  2022, \mn@doi [\apjl] {10.3847/2041-8213/ac7f44}, \href
  {https://ui.adsabs.harvard.edu/abs/2022ApJ...934L..11V} {934, L11}

\bibitem[\protect\citeauthoryear{{Weingartner} \& {Draine}}{{Weingartner} \&
  {Draine}}{2001}]{Weingartner2001}
{Weingartner} J.~C.,  {Draine} B.~T.,  2001, \mn@doi [\apj] {10.1086/318651},
  \href {https://ui.adsabs.harvard.edu/abs/2001ApJ...548..296W} {548, 296}

\bibitem[\protect\citeauthoryear{{Wilking} et~al.}{{Wilking}
  et~al.}{2005}]{Wilking2005}
{Wilking} B.~A.,  et~al., 2005, \mn@doi [\aj] {10.1086/432758}, \href
  {https://ui.adsabs.harvard.edu/abs/2005AJ....130.1733W} {130, 1733}

\bibitem[\protect\citeauthoryear{{Zhang} et~al.}{{Zhang}
  et~al.}{2018}]{Zhang2018}
{Zhang} S.,  et~al., 2018, \mn@doi [\apjl] {10.3847/2041-8213/aaf744}, \href
  {https://ui.adsabs.harvard.edu/abs/2018ApJ...869L..47Z} {869, L47}

\bibitem[\protect\citeauthoryear{{Zhu}}{{Zhu}}{2015}]{Zhu2015}
{Zhu} Z.,  2015, \mn@doi [\apj] {10.1088/0004-637X/799/1/16}, \href
  {https://ui.adsabs.harvard.edu/abs/2015ApJ...799...16Z} {799, 16}

\makeatother
\end{thebibliography}
\bibliographystyle{mnras}

\bsp
\label{lastpage}

\end{document}